\documentclass[12pt]{iopart}

\usepackage{graphicx,dsfont}
\usepackage{color}
\usepackage{iopams}
\usepackage{bbm}
\definecolor{red}{rgb}{1,0,0}

\begin{document}
\title{Eigenvalue Density of the Doubly Correlated Wishart Model: Exact Results
}
\author{Daniel Waltner, Tim Wirtz, Thomas Guhr}
\address{Fakult\"at f\"ur Physik, Universit\"at Duisburg-Essen, Lotharstra\ss e 1, 47048 Duisburg, Germany}

\date{}

\begin{abstract}
Data sets collected at different times and different observing points can possess correlations at different times {\it and} at different positions. The doubly correlated Wishart model takes both into
account. We calculate the
eigenvalue density of the Wishart correlation matrices using supersymmetry. In the complex case we obtain a new closed form expression which we compare to previous results in the literature. In the 
more relevant and much more  complicated real
case we derive an expression for the density in terms of a fourfold integral. Finally, we calculate the density in the limit of large correlation matrices.
\end{abstract}
\maketitle

\section{Introduction}
The concept of random matrix theory (RMT), \textit{i.e.}\ the idea to replace matrices appearing in the description of systems by matrices only constraint by the symmetries of the underlying system
was originally put forward by Wishart in the context of biostatistics \cite{Wish}. Later on, random matrices modelling dynamical systems were introduced by Wigner \cite{Wig}.
Nowadays, RMT is an important tool in quantum chaos \cite{Stoeck}, disordered systems \cite{Datta}, quantum chromodynamics \cite{Verba} and the theory of wireless communications \cite{wire}. An
improvement concerning the computational methods was achieved by supersymmetry by introducing combinations of integrals with respect to commuting (bosonic) and anticommuting (fermionic) variables.

Here, we are concerned with the analysis of time and position series. A data set is collected by measuring $p$ different observables at $n$ different times. Examples are the electrical activity
measured along the scalp at $p$ different places within an
electroencephalography, the temperature or the water level of a river at $p$ different places, the prices
of $p$ different stocks each measured at $n$ different times. 
Such analyses of spatio-temporal correlations were performed for example for climate change detection \cite{Ribes}, criminal offense reports \cite{Toole} or macroeconomic data \cite{Snarska}.   
The data set forms the empirical data matrix. The rows represent time series, \textit{i.e.}\
the measurements at one of the $p$ points at different times and the columns represent position series, \textit{i.e.}\ the measurements at a fixed time for all $p$ positions. We are now interested
in correlations between the time and position series. To study generic features we make an ensemble approach and model these correlations by the doubly correlated Wishart model \cite{Simon, SimonI,Burda,McKay} that depends on two empirical correlation matrices, one for the correlations
of the time series and another one for those of the position series. We thus aim at deriving generic statistical properties. We consider complex and real data. At present, it is not clear to 
us how to obtain these results by methods other than supersymmetry.

This article is organized as follows: In Sec.\ \ref{sect2} we sketch the salient features of the doubly correlated Wishart model. In Sec.\ \ref{sect2a} we
show how to obtain the eigenvalue density of the Wishart correlation matrices by supersymmetry. As the further analysis is quite different for the complex and the real case, we treat them separately.

\section{Doubly Correlated Wishart Model}\label{sect2}

We consider a data matrix $M$ resulting from data of $p$ observables collected at $n$ time steps. The rows are the time and the columns the position series. The data set can be real as well as
complex. As data obtained in everyday life are
usually real, the latter case is however the more relevant one. The correlations between the rows of this matrix, \textit{i.e.}\ the time series, are after
normalization to zero mean and unit variance measured by the $p\times p$ matrix
\begin{equation}\label{eq100000}
 C=\frac{1}{n}MM^\dagger
\end{equation}
with $C_{jl}$ describing the correlation between the $j$th and $l$th time series. Analogously, the $n\times n$ matrix
\begin{equation}\label{eq100001}
 D=\frac{1}{p}M^\dagger M
\end{equation}
determines correlations between position series. 
Because in general $n\neq p$, either $C$ or $D$ as introduced in Eqs.\ (\ref{eq100000},\ref{eq100001}) does not have full rank. Since these are used to model the correlation structure, it 
is reasonable to assume that they are independently measured in a way such that both matrices have full rank. We will refer to these as empirical correlation matrices. 

In order to gain information about the statistical properties of the data matrices $M$ they are replaced by random matrices $W$ which can be viewed as model data matrices. In the doubly correlated
Wishart model
%
the entries of $W$ are Gaussian distributed according to
\begin{equation}\label{eq1}
P_\beta(W|C,D)=K_\beta\exp\left(-\frac{\beta}{2}\tr D^{-1}W^\dagger C^{-1}W\right)
\end{equation}
with the symmetry index $\beta$ taking the values $\beta=1$ for $W \in \mathds{R}^{p\times n}$ and $\beta=2$ for $W \in \mathds{C}^{p\times n}$. The  real symmetric (complex hermitean) $p\times p$ matrix $C$ and the
real symmetric (complex hermitean) $n\times n$ matrix $D$ for $\beta=1$ ($\beta=2$) are the empirical correlation matrices and are model input. The measure for calculating averages with respect to
$W$ is given by $P_\beta(W|C,D)d[W]$ where
\begin{equation}
d[W]=\left\{\begin{array}{cc}\prod_{j=1}^p\prod_{k=1}^n dW_{jk}&\rm{for}\,\beta=1\\
\prod_{j=1}^p\prod_{k=1}^nd{\rm{Re}}W_{jk}d{\rm{Im}}W_{jk}&\rm{for}\,\beta=2
\end{array}\right..
\end{equation}
It fulfills the invariance
\begin{equation}
\label{invari}
P_\beta(W|C,D)d[W]=P_\beta(UWV|UCU^\dagger,V^\dagger DV)d[UWV]
\end{equation}
with arbitrary orthogonal ($\beta=1$) or unitary ($\beta=2$) $p\times p$ matrix $U$ and $n\times n$ matrix $V$. The constant $K_\beta$ in Eq.\ (\ref{eq1}) is determined by the
condition that $P_\beta(W|C,D) d[W]$ should be normalized,
\begin{equation}
K_\beta=(2\pi/\beta)^{-\beta np/2}\det\left(D^{-1}\otimes C^{-1}\right)^{\beta/2}
\end{equation}
where $D^{-1}\otimes C^{-1}$ denotes the tensor product of the matrices $D^{-1}$ and $C^{-1}$.

The matrix $WW^\dagger/n$ is the model Wishart position correlation matrix while $W^\dagger W/p$ is the Wishart time correlation matrix. The choice of the distribution (\ref{eq1}) ensures that their
averages coincide with the corresponding empirical correlation matrices,
\begin{equation}
\fl\quad\quad\quad \frac{1}{n}\int d[W]P_\beta(W|C,D)WW^\dagger=C,\hspace*{1cm}\frac{1}{p}\int d[W]P_\beta(W|C,D)W^\dagger W=D.
\end{equation}
We are interested in the density $S_\beta(x)$ of eigenvalues $\lambda_j$ of $WW^\dagger$ defined by
\begin{eqnarray}\label{dens}
S_\beta(x)=\frac{1}{p}\int d[W]P_\beta(W|C,D)\sum_{j=1}^p\delta(x-\lambda_j)\nonumber\\=\frac{1}{\pi p}\lim_{\epsilon\rightarrow 0^+}{\rm Im}\int d[W]P_\beta(W|C,D)
\tr\frac{\mathds{1}_p}{(x-i\epsilon)\mathds{1}_p-WW^\dagger}.
\end{eqnarray}
Analogously, the density $\tilde{S}_\beta(x)$ of the eigenvalues $\gamma_j$ of $W^\dagger W$ is given by 
\begin{eqnarray}\label{dens50}
{\tilde S}_\beta(x)=\frac{1}{n}\int d[W]P_\beta(W|C,D)\sum_{j=1}^n\delta(x-\gamma_j)\nonumber\\=\frac{1}{\pi n}\lim_{\epsilon\rightarrow 0^+}{\rm Im}\int d[W]P_\beta(W|C,D)
\tr\frac{\mathds{1}_n}{(x-i\epsilon)\mathds{1}_n-W^\dagger W}.
\end{eqnarray}
The expressions in the second line of Eqs.\ (\ref{dens},\ref{dens50}) are related because of
\begin{equation}
\tr\frac{\mathds{1}_n}{(x-i\epsilon)\mathds{1}_n-W^\dagger W}=\tr\frac{\mathds{1}_p}{(x-i\epsilon)\mathds{1}_p-WW^\dagger}-\frac{p-n}{x-i\epsilon}.
\end{equation}
Thus we have
\begin{equation}
\tilde{S}_\beta(x)=\frac{p}{n}S_\beta(x)+\left(1-\frac{p}{n}\right)\delta(x). 
\end{equation}
If we are interested instead in the eigenvalue density of the matrices $WW^\dagger/n$ and $W^\dagger W/p$ in analogy to Eqs.\ (\ref{eq100000},\ref{eq100001}), we obtain the latter using the properties of delta distributions
to be given by $nS_\beta(nx)$ and $p{\tilde S}_\beta(px)$, respectively.

Due to the invariance relation (\ref{invari}) we can assume without loss of generality the matrices $C$ and $D$ to be diagonal. We will denote these diagonal matrices
in the following by $\Lambda={\rm{diag}}(\Lambda_1,\dots,\Lambda_p)$ and $\Gamma={\rm{diag}}(\Gamma_1,\ldots,\Gamma_n)$ referred to as empirical ones.
\section{Supersymmetry Approach}\label{sect2a}
We now calculate $S_\beta(x)$ using supersymmetry. The calculation extends the one in Ref.\ \cite{Rech12} where the one sided correlated Wishart model, \textit{i.e.}\ $D=\mathds{1}_n$, was studied.
Although the basic steps are standard, we present them here to make this presentation self-contained. We introduce the generating function
\begin{equation}\label{erzeug}
Z_\beta(x_0,x_1)= \int d[W]P_\beta(W|\Lambda,\Gamma)\frac{\det(x_1\mathds{1}_p-WW^\dagger)}{\det(x_0\mathds{1}_p-WW^\dagger)},
\end{equation}
which yields the eigenvalue density (\ref{dens}) by the relation
\begin{equation}\label{specden}
S_\beta(x)=\frac{1}{2\pi ip}\frac{\partial}{\partial J}\left.\lim_{\epsilon\rightarrow 0^+}\left[Z_\beta(x-i\epsilon,x+J)-Z_\beta(x+i\epsilon,x+J)\right]\right|_{J=0}.
\end{equation}
The ratio of determinants in the generating function is expressed as a supermatrix integral
\begin{equation}\label{super}
\frac{\det(x_1\mathds{1}_p-WW^\dagger)}{\det(x_0\mathds{1}_{p}-WW^\dagger)}=\int d[A]{\exp}\left[{\frac{i\beta}{2}{\rm{str}}(XA^\dagger A-A^\dagger WW^\dagger A)}\right]
\end{equation}
with $X={\rm diag}(x_0\mathds{1}_{2/\beta},x_1\mathds{1}_{2/\beta})$ and the supermatrix $A$
\begin{eqnarray}
\begin{array}{ccc}A=[u\,\,\,v\,\,\,\zeta^*\,\,\,\zeta],&A^\dagger=\left[\begin{array}{c} u^T\\ v^T\\ \zeta^T\\ -\zeta^{*T}\end{array}\right]&{\rm for}\,\beta=1,\\A=[z^*\,\,\,\zeta^*],&A^\dagger=\left[
\begin{array}{c} z^T\\ \zeta^T\end{array}
\right]&{\rm for}\,\beta=2\end{array}
\end{eqnarray}
containing the vectors $u,v\in\mathds{R}^p$, $z\in\mathds{C}^p$ and the $p$ component vectors $\zeta,\zeta^*$ containing anticommuting variables. The measure $d[A]$ is defined by
\begin{equation}
d[A]=\left\{\begin{array}{cc}(2\pi)^{-p}\prod_{j=1}^p
du_jdv_jd\zeta_j^*d\zeta_j&{\rm for}\,\beta=1\\ \pi^{-p}\prod_{j=1}^pd{\rm Re}z_jd{\rm Im}z_jd\zeta_j^*d\zeta_j&{\rm for}\,\beta=2\end{array}\right..
\end{equation}
Using the representation (\ref{super}) we obtain for the generating function (\ref{erzeug})
\begin{eqnarray}
Z_\beta(x_0,x_1)&=&K_\beta\int d[A]{\exp}{\left[i\frac{\beta}{2}{\rm str}(XA^\dagger A)\right]}\\ &&\times\int d[\vec{W}]{\exp}{\left\{-\frac{\beta}{2}{\rm tr}\left[\vec{W}(\Gamma^{-1}\otimes \Lambda^{-1}+i\mathds{1}_n
\otimes AA^\dagger)\vec{W}\right]\right\}}\nonumber
\end{eqnarray}
with the supertrace denoted by str and the $(n\cdot p)$ component vector $\vec{W}$ containing the elements of the matrix $W$. Performing the Gaussian integrals with respect to the components of 
$\vec{W}$ yields
\begin{equation}\label{genera}
\fl \quad \quad \quad Z_\beta(x_0,x_1)=\int d[A]{\exp}\left[{i\frac{\beta}{2}{\rm str}(XA^\dagger A)}\right]{\rm det}^{-\beta/2}(\mathds{1}_n\otimes\mathds{1}_p+i\Gamma\otimes \Lambda AA^\dagger).
\end{equation}
Because of the dyadic structure the determinant in the last equation equals the superdeterminant ${\rm sdet}(\mathds{1}_n\otimes\mathds{1}_{4/\beta}+i\Gamma\otimes A^\dagger \Lambda A)$ \cite{Rech12}. To perform the
superintegrals with respect to $A$ we employ the Fourier representation of the latter superdeterminant
\begin{equation}\label{Ihs}
\fl\quad\quad\quad {\rm sdet}^{-\beta/2}(\mathds{1}_n\otimes\mathds{1}_{4/\beta}+i\Gamma\otimes A^\dagger \Lambda A)=\frac{\beta^2}{4}\int d[\rho]I_\beta(\rho|\Gamma){\exp}\left[{-i\frac{\beta}{2}{\rm str}A^\dagger \Lambda A\rho}\right]
\end{equation}
with the supermatrix $\rho$ possessing the parametrization
\begin{equation}\label{rho}
\rho=\left[\begin{array}{cc}\rho_0&\chi\\\tilde{\chi}&i\rho_1
\mathds{1}_{2/\beta}\end{array}\right],
\end{equation}
where all the entries are $2/\beta\times2/\beta$ square matrices. The matrices on the diagonal contain commuting variables, $\rho_1$ is a real number, $\rho_0$ is  a real number  for $\beta=2$
and a real symmetric $2\times 2$ matrix for $\beta=1$ that we parametrize as
\begin{equation}\label{rho0}
\rho_0=\left[\begin{array}{cc}\rho_{00}&\rho_{01}\\ \rho_{01}&\rho_{11}\end{array}\right].
\end{equation}
The $i$ in front of $\rho_1$ in Eq.\ (\ref{rho}) was introduced to ensure convergence of the integral with respect to the matrix $\rho$ \cite{Rech12}.
The matrices $\chi$ and $\tilde{\chi}$ only contain Grassmann variables, for $\beta=2$ $\chi$ and $\tilde{\chi}=\chi^*$ are both scalar, for $\beta=1$ they possess the following matrix parametrizations
\begin{equation}\label{Grassma}
\chi=\left[\begin{array}{cc}\eta&\eta^*\\ \xi&\xi^*\end{array}\right],\hspace*{2cm}\tilde{\chi}=\left[\begin{array}{cc}\eta^*&\xi^*\\ -\eta&-\xi\end{array}\right].
\end{equation}
In contrast to Ref.\ \cite{Rech12} we do not find an Ingham Siegel type of integral \cite{IngSie} here but a generalization thereof
\begin{eqnarray}\label{Ingha}
I_{\beta}(\rho|\Gamma)&=&\int d[\sigma]{\rm sdet}^{-\beta/2}(\mathds{1}_n\otimes\mathds{1}_{4/\beta}+i\Gamma\otimes\sigma){\exp}\left[{i\frac{\beta}{2}{\rm str}(\rho\sigma)}\right]\nonumber\\&=&
\int d[\sigma]\prod_{j=1}^n{\rm sdet}^{-\beta/2}(\mathds{1}_{4/\beta}+i\Gamma_j\sigma){\exp}\left[{i\frac{\beta}{2}{\rm str}(\rho\sigma)}\right].
\end{eqnarray}
Due to the dependence on the entries $\Gamma_j$ of $\Gamma$, it encodes all information about the time correlations. The matrix $\sigma$ possesses the same form as $\rho$. The commuting variables
of $\sigma$ we denote analogously to Eqs.\ (\ref{rho}, \ref{rho0}).
The off-diagonal blocks in $\sigma$ we denote by $\omega$ and $\tilde{\omega}$ and parametrize them for $\beta=1$ by
\begin{equation}
\omega=\left[\begin{array}{cc}\alpha&\alpha^*\\ \beta&\beta^*\end{array}\right]\hspace*{2cm}\tilde{\omega}=\left[\begin{array}{cc}\alpha^*&\beta^*\\ -\alpha&-\beta\end{array}\right].
\end{equation}
The generating function in Eq.\ (\ref{genera}) can be expressed in terms of an integral with respect to a $p\cdot 4/\beta$ component supervector $\vec{A}$ 
\begin{equation}
 Z_\beta(x_0,x_1)=\frac{1}{4}\int d[\rho]I_\beta(\rho|\Gamma)\int d[\vec{A}]\exp\left[i\frac{\beta}{2}\vec{A}^\dagger\left(\mathds{1}_p\otimes X-\Gamma\otimes\rho\right)\vec{A}\right]
\end{equation}
Performing the Gaussian integral with respect to the components of $\vec{A}$ yields the following matrix integral representation
of the generating function in superspace
\begin{eqnarray}\label{eqgen}
Z_\beta(x_0,x_1)&=\frac{1}{4}
\int d[\rho]I_\beta(\rho|\Gamma)\,{\rm sdet}^{-\beta/2}(\mathds{1}_p\otimes X-\Lambda\otimes\rho)\nonumber\\&=\frac{1}{4}\int d[\rho]\,I_\beta(\rho|\Gamma)\prod_{j=1}^p{\rm sdet}^{-\beta/2}
(X-\Lambda_j\rho).
\end{eqnarray}
Thus we have reduced the number of integrals from $\beta\cdot n\cdot p$ integrals with respect to commuting variables in Eq.\ (\ref{erzeug}) to $8/\beta$ integrals with respect to commuting and $8/\beta$ integrals with respect to anticommuting variables.

Using superbosonization \cite{Zirnba} instead of the Ingham Siegel integral as in Ref.\ \cite{Rech12} to compute
$Z(x_0,x_1)$ seems not applicable here to us since the integrand in Eq.\ (\ref{genera}) is not invariant under the transformation $A\mapsto UA$ with $U$ a $p\times p$ unitary
(orthogonal) matrix for $\beta=2$ ($\beta=1$). It would be interesting to understand from a group theoretical point of view in the setting of superbosonization the consequences of the empirical
eigenvalues of the time correlation matrix.
\section{Explicit Expressions for the Density in the Complex Case}\label{sect3}
As the calculations following are substantially different in the cases $\beta=2$ and $\beta=1$, we treat them separately and start in this section with the unitary case $\beta=2$.
We derive an explicit expression for the density in Sec.\ 4.1 and compare it in Sec.\ 4.2 with a previous result in the literature \cite{Simon} and with Monte Carlo simulations in Sec.\ 4.3.
\subsection{Supersymmetric Expression for the Density}
With the parametrization described in Eq.\ (\ref{rho}) we expand $I_2(\rho|\Gamma)$  in the Grassmann
variables $\omega$, $\omega^*$, $\chi$, $\chi^*$ and obtain
\begin{eqnarray}\label{inghamm}
\fl I_2(\rho|\Gamma)=\int d[\sigma]\prod_{j=1}^n\frac{1-\Gamma_j\sigma_1}{1+i\Gamma_j\sigma_0}\left[1-\sum_{l=1}^n\frac{\Gamma_l^2\omega\omega^*}{(1-\Gamma_l\sigma_1)
(1+i\Gamma_l\sigma_0)}\right](1+\omega\omega^*\chi\chi^*){\rm e}^{i(\sigma_0\rho_0+\sigma_1\rho_1)}.\nonumber\\
\end{eqnarray}
The integrals with respect to  the Grassmann variables $\omega$, $\omega^*$ can be performed directly.
Replacing $\sigma_1$ in the preexponential factor in (\ref{inghamm}) by $-i\partial/\partial\rho_1$, the $\sigma_1$ integral yields $\delta(\rho_1)$. The
$\sigma_0$ integral is performed with help of the residue theorem, the integration contour is closed in the upper complex half plane for $\rho_0>0$ and in the lower half
plane for $\rho_0<0$. Since it does not have poles for $\rho_0<0$ this part is zero, whereas for $\rho_0>0$ we find
\begin{eqnarray}
&\fl I_2(\rho|\Gamma)=\frac{4\pi \Theta(\rho_0)}{(-1)^{n}\prod_{j=1}^n\Gamma_j}\sum_{k=1}^n\left[\frac{\prod_{j=1}^n
\left(1+i\Gamma_j\frac{\partial}{\partial\rho_1}\right)}{\Delta(\Gamma,\Gamma_k)}\chi\chi^*{\rm e}^{-\rho_0/\Gamma_k}+\prod_{j=1,\atop j\neq
k}^n\left(1+i\Gamma_j\frac{\partial}{\partial\rho_1}\right)\right.\nonumber\\&\fl\left.\Gamma_k\left(\sum_{l=1,\atop l\neq k}^n\frac{{\rm e}^{-\rho_0/
\Gamma_l}}
{\Delta(\Gamma,\Gamma_l)\left(\frac{1}
{\Gamma_l}-\frac{1}{\Gamma_k}\right)}-\frac{\rho_0{\rm e}^{-\rho_0/\Gamma_k}}{\Delta(\Gamma,\Gamma_k)}-\sum_{l=1,\atop l\neq k}
\frac{{\rm e}^{-\rho_0/\Gamma_k}}{\Delta(\Gamma,\Gamma_l)\left(\frac{1}
{\Gamma_k}-\frac{1}{\Gamma_l}\right)}\right)\right]\delta(\rho_1)
\end{eqnarray}
with $\Theta(\rho_0)$ the Heaviside theta function. We defined
\begin{equation}
\Delta(\Gamma,\Gamma_l)=\prod_{j=1,\atop j\neq l}^n\left(\frac{1}{\Gamma_l}-\frac{1}{\Gamma_j}\right).
\end{equation}
To compute $Z_2(x_0,x_1)$ by means of Eq.\ (\ref{eqgen}), we expand ${\rm sdet}^{-1}(X-\Lambda_j\rho)$ in terms of the Grassmann variables $\chi$, $\chi^*$
\begin{eqnarray}\label{eq999}
\fl{\rm sdet}^{-1}(X-\Lambda_j\rho)=\prod_{j=1}^p\frac{x_1-i\Lambda_j\rho_1}
{x_0-\Lambda_j\rho_0}\left(1+\sum_{k=1}^p\frac{\Lambda_k^2\chi\chi^*}{(x_0-\Lambda_k\rho_0)(x_1-i\Lambda_k\rho_1)}\right).
\end{eqnarray}
Inserting $I_2(\rho|\Gamma)$ as well as the last expression into Eq.\ (\ref{eqgen}), we can perform the integrals with respect to the anticommuting variables,
the $\rho_1$ integral after partial integrations using the identity $\int_{-\infty}^\infty d\rho_1\,\rho_1^n\delta(\rho_1)=\delta_{0n}$ and finally the
$\rho_0$ integral by using the identity
\begin{equation}\label{delta00}
\lim_{\epsilon\rightarrow 0}\frac{1}{\rho_0-x_0/\Lambda_j+i\epsilon}=-i\pi\delta\left(\rho_0-\frac{x_0}{\Lambda_j}\right)+P\left(
\frac{1}{\rho_0-x_0/\Lambda_j}\right)
\end{equation}
as well as the derivative of the latter relation with respect to $\rho_0$. Eventually, we are interested in the eigenvalue density, \textit{i.e.}\ the imaginary part of $Z(x_0,x_1)$ for $\epsilon
\rightarrow 0$. Thus,
the $\rho_0$ integral can be always performed by the delta function in
Eq.\ (\ref{delta00}) because products of principal value terms only do not contribute.
We then obtain for the spectral density via the relation (\ref{specden})
\begin{eqnarray}
\label{dens1}
&\fl S_2(x)=\frac{1}{p(-1)^{p+n}\left(\prod_{j=1}^n\Gamma_j\right)\left(\prod_{j=1}^p\Lambda_j\right)}\left\{\sum_{k=1}^n\sum_{l=1}^p
\frac{{\rm e}^{-x/(\Gamma_k\Lambda_l)}}{x^{p-1}\Delta(\Lambda,\Lambda_l)\Delta(\Gamma,\Gamma_k)}\sum_{u=0}^{p-1}E_u^n(\Gamma)u!\right.\nonumber\\&\fl\left.\times E_u^p
(\Lambda)(p-u)(-1)^ux^{p-u-1}+\sum_{k=1}^n\sum_{m=1}^p\left[\frac{\Lambda_m\Gamma_k}{x^{p-1}\Delta(\Lambda,\Lambda_m)}\left(\sum_{l=1,\atop l\neq k}^n\frac{
{\rm e}^{-x/(\Lambda_m\Gamma_l)}}{\Delta(\Gamma,\Gamma_l)\left(1-\frac{\Gamma_l}{\Gamma_k}\right)}\right.\right.\right.\nonumber\\&\fl\left.\left.\left.-\frac{\left(\frac{x}{\Lambda_m\Gamma_k}
-1\right){\rm e}^{-x/(\Lambda_m\Gamma_k)}}{
\Delta(\Gamma,\Gamma_k)}-\sum_{l=1,\atop l\neq k}^n\frac{{\rm e}^{-x/(\Lambda_m\Gamma_k)}}{\Delta(\Gamma,\Gamma_k)\left(1-
\frac{\Gamma_k}{\Gamma_l}\right)}\right)+\sum_{q=1,\atop q\neq m}^p\frac{\Lambda_m\Gamma_k}{x^p\Delta(\Lambda,\Lambda_m)\left(\frac{1}{\Lambda_m}-\frac{1}{\Lambda_q}\right)}\right.\right.\nonumber\\&\fl
\left.\left.\left(\sum_{l=1,\atop l\neq k}^n\frac{
{\rm e}^{-x/(\Lambda_m\Gamma_l)}}{\Delta(\Gamma,\Gamma_l)\left(\frac{1}{\Gamma_l}-\frac{1}{\Gamma_k}\right)}-\frac{x
{\rm e}^{-x/(\Lambda_m\Gamma_k)}}{\Lambda_m
\Delta(\Gamma,\Gamma_k)}-\sum_{l=1,\atop l\neq k}^n\frac{{\rm e}^{-x/(\Lambda_m\Gamma_k)}}{\Delta(\Gamma,\Gamma_k)\left(\frac{1}{\Gamma_k}-
\frac{1}{\Gamma_l}\right)}\right)+\right.\right.\nonumber\\&\fl\left.\left.\sum_{q=1,\atop q\neq m}^p\frac{\Lambda_m\Gamma_k}{x^p\Delta(\Lambda,\Lambda_q)\left(\frac{1}{\Lambda_m}-
\frac{1}{\Lambda_q}\right)}\left(\sum_{l=1,\atop l\neq k}^n\frac{
{\rm e}^{-x/(\Lambda_q\Gamma_l)}}{\Delta(\Gamma,\Gamma_l)\left(\frac{1}{\Gamma_l}-\frac{1}{\Gamma_k}\right)}-\frac{x
{\rm e}^{-x/(\Lambda_q\Gamma_k)}}{\Lambda_q
\Delta(\Gamma,\Gamma_k)}\right.\right.\right.\nonumber\\&\fl\left.
\left.\left.-\sum_{l=1,\atop l\neq k}^n\frac{{\rm e}^{-x/(\Lambda_q\Gamma_k)}}{\Delta(\Gamma,\Gamma_k)\left(\frac{1}{\Gamma_k}-
\frac{1}{\Gamma_l}\right)}\right)\right]\sum_{u=0}^{p-2}E_u^n(\Gamma^{\hat{k}})
E_u^p(\Lambda^{\hat{m}})u!(-1)^u(p-u-1)x^{p-u-2}\right\},
\end{eqnarray}
where we defined the elementary symmetric functions
\begin{eqnarray}
E_k^p(\Lambda)=\hspace*{-6mm}\sum_{1\leq j_1<\ldots<j_k\leq p}\Lambda_{j_1}\ldots\Lambda_{j_k}\hspace*{1.3cm}
E_k^n(\Gamma)=\hspace*{-6mm}\sum_{1\leq j_1<\ldots<j_k\leq n}\Gamma_{j_1}\ldots\Gamma_{j_k}.
\end{eqnarray}
If the $l$th eigenvalue is excluded in $E_k^p(\Lambda)$ or $E_k^n(\Gamma)$ we denote this quantity by  $E_k^p(\Lambda^{\hat l})$, $E_k^n(\Gamma^{\hat l})$, respectively. These
quantities are by definition equal to one for $k=0$.
\subsection{Comparison with Previous Results}
We compared our result for $S_2(x)$ in Eq.\ (\ref{dens1}) with the one obtained by the character expansion method in Ref.\ \cite{Simon}. There the eigenvalue density was calculated using instead of the distribution (\ref{eq1}) a distribution
proportional to $\exp\left(-\tr BM^\dagger AM\right)$ with $M\in\mathds{C}^{n\times p}$ and the correlation matrices $A$ and $B$. For comparison, we thus need to replace the eigenvalues $a_i$ and $b_i$ of the matrices $A$ and $B$ by the eigenvalues
$\Gamma_i^{-1}$ and $\Lambda_i^{-1}$ of the matrices $D^{-1}$ and $C^{-1}$, respectively.

For completeness we first give here the result in Ref.\ \cite{Simon}.
Defining the Vandermonde determinant
\begin{equation}
 \Delta_V(x)=\prod_{1\leq i<j\leq V}(x_j-x_i)
\end{equation}
and the expression
\begin{equation}
 g(x;\alpha,z)=x^{n-\alpha}(\alpha-1)!\sum_{m=0}^{\alpha-1}\frac{(-zx)^m}{m!},
\end{equation}
the density is given in Eq.\ (16) in \cite{Simon} as a sum of two determinants
\begin{eqnarray}
\label{refden}
\fl S_2(x)=\Delta_n(a)\Delta_p(b)(-x)^{p(p-1)/2}\prod_{j=1}^{n-1}j^j\prod_{j=1}^{n-p-1}(n-j)^{n-j-p}\sum_{k=1}^n\left(\det \tilde{K}^{(k)}+\sum_{m=1,\atop m\neq k}^n\det T^{(k,m)}\right)\nonumber\\
\end{eqnarray}
with the $n\times n$ matrices $\tilde{K}^{(k)}$ and $T^{(k,m)}$. For $i\neq k$ the elements of $\tilde{K}^{(k)}$ are given by $\tilde{K}_{ij}^{(k)}=g(a_ib_j;n,x)$ for $j\leq p$ and $\tilde{K}_{ij}^{(k)}=
a_i^{j-1}$ for $j>p$ and for $i=k$ $\tilde{K}_{kj}^{(k)}=\left[a_kb_j+p(p-1)/(2x)\right]\exp\left(-a_kb_j x\right)(n-1)!$ for $j\leq p$ and $\tilde{K}_{kj}^{(k)}=0$ $j>p$. The elements of
$T^{(k,m)}$ are given for $i\neq m$, $i\neq k$ by $T^{(km)}_{ij}=g(a_ib_j;n,x)$ for $j\leq p$ and by $T^{(km)}_{ij}=a_i^{j-1}$ for $j>p$, for $i\neq m$, $i=k$ by $T^{(km)}_{kj}=\exp\left(-a_kb_jx
\right)(n-1)!$ for $j\leq p$ and $T^{(km)}_{kj}=0$ for $j>p$ and for $i= m$ by $T^{(km)}_{mj}=(n-1)g(a_mb_j;n-1,x)$ for $j\leq p$ and $T^{(km)}_{mj}=0$ for $j>p$.  We note that there is an error in
$T^{(km)}_{mj}$ in Ref.\ \cite{Simon} where an additional factor $a_mb_j$ is contained in the corresponding expression for $j\leq p$.

\begin{figure}
\begin{center}
\includegraphics[width=7.2cm]{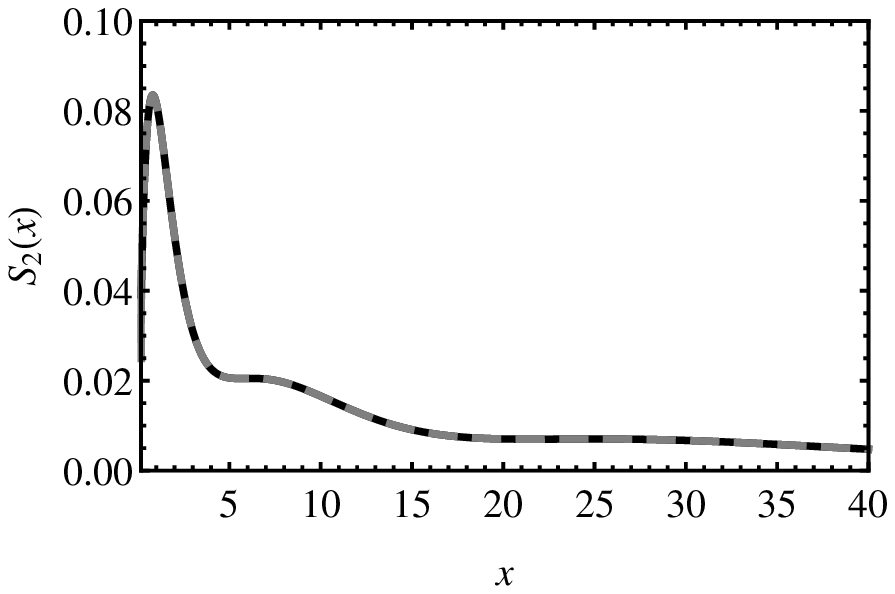}\raisebox{-1.5mm}{\includegraphics[width=8.2cm]{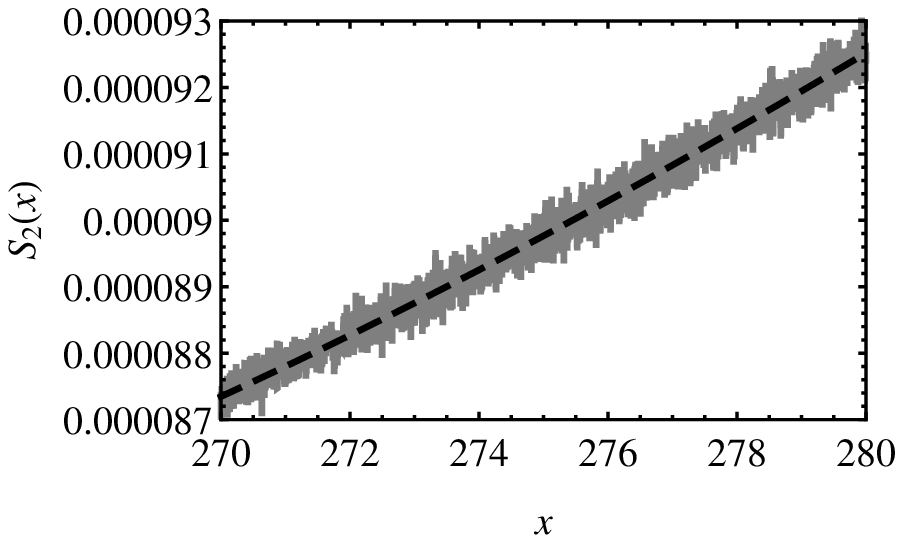}}
\caption{We show plots of the eigenvalue density for $\Lambda_j=5.9,\, 1.1,\, 4.2,\, 2,\, 50$ and $\Gamma_j=0.3,\, 2.7,\, 2.5,\, 2.8,\, 5.6,\, 1$ resulting from the expressions (\ref{dens1}) (black dashed curve) and
(\ref{refden}) (full gray curve). On the right hand side we resolve fluctuations for large x values resulting in case of a plot of (\ref{refden}) whereas Eq.\ (\ref{dens1}) yields the
smooth curve depicted.}
\label{fig2}
\end{center}
\end{figure}

Plotting the densities (\ref{dens1}) and (\ref{refden}) for certain $\Lambda_i$ and $\Gamma_i$, we find agreement between the two expressions, see the plot on the left hand side of Fig.\ \ref{fig2}, however our expression seems more suitable for numerical
evaluation as the expression (\ref{refden}) often shows for large $x$ values huge fluctuations. This can be observed in the plot on the right hand side in Fig.\ \ref{fig2}, where we compare both expressions for $\Lambda_j=5.9,\, 1.1,\,
4.2,\, 2,\, 50$ and $\Gamma_j=0.3,\, 2.7,\, 2.5,\, 2.8,\, 5.6,\, 1$ in the range $x\in[270,280]$.

\subsection{Comparison with Monte Carlo Simulation}
\begin{figure}
\begin{center}
\includegraphics[width=8cm]{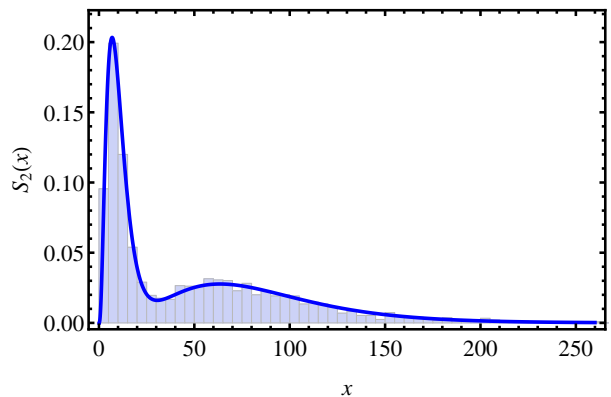}
\caption{The density $S_2(x)$ resulting from Eq.\ (\ref{dens1}) is shown for $\Lambda_j=5.9,\, 1.1$ and $\Gamma_j=0.3,\, 2.7,\, 2.5,\, 2.8,\, 5.6$ (full blue curve) and compared with a Monte Carlo simulation shown as histogram.}
\label{fig1}
\end{center}
\end{figure}

An alternative to calculate the eigenvalue density of $WW^\dagger$ with the distribution (\ref{eq1}) is to consider the eigenvalue density of $\sqrt{C}WDW^\dagger\sqrt{C}$ averaged with the
Gaussian measure proportional to $\exp\left(-\tr W^\dagger W\right)$.
We follow this approach here, consider $100000$ realizations of matrices $W$ and show in Fig.\ \ref{fig1} a comparison between our expression for $\Lambda_j=5.9,\, 1.1$ and $\Gamma_j=0.3,\, 2.7,\, 2.5,\, 2.8,\, 5.6$ and a corresponding Monte Carlo
simulation and find excellent agreement.

\section{Real Case}\label{sect3a}
In principle, we make here the same steps as in the previous Sec.\ \ref{sect3}. However,
the analytical calculation is now much more complicated due to the larger number of integration variables and square root singularities. For $I_1(\rho|\Gamma)$ in  Eq.\ (\ref{Ingha}) we get here after performing the
integrals with respect to the Grassmann variables contained in the matrix $\sigma$
\begin{eqnarray}\label{I_1}
&\fl I_1(\rho|\Gamma)=\frac{1}{4\pi^2}\int d[\sigma_0]d\sigma_1{\rm e}^{i\tr(\sigma_0\rho_0)/2+i\sigma_1\rho_1}\prod_{j=1}^n\frac{1}{\det^{1/2}\left(\mathds{1}_2+i\Gamma_j\sigma_0\right)}\left[
 \eta^*\eta\xi^*\xi\prod_{j=1}^n(1-\Gamma_j\sigma_1)\right.\nonumber\\&\fl\left.+\sum_{k=1}^n\frac{\Gamma_k^2\left[\xi^*\xi(1+i\Gamma_k\sigma_{11})+
 \eta^*\eta(1+i\Gamma_k\sigma_{00})-i\Gamma_k\sigma_{10}(\eta\xi^*-\eta^*\xi)\right]}
 {\det(\mathds{1}_2+i\Gamma_k\sigma_0)}
\prod_{j=1,\atop j\neq k}^n(1-\Gamma_j\sigma_1) \right.\nonumber\\&\fl\left.+\sum_{k,l=1,\atop k\neq l}^n\frac{\Gamma_k^2\Gamma_l^2\left[(1+i\Gamma_k\sigma_{11})(1+i\Gamma_l\sigma_{00})+\Gamma_k\Gamma_l\sigma_{10}^2\right]}
{\det(\mathds{1}_2+i\Gamma_k\sigma_0)\det(\mathds{1}_2+i\Gamma_l\sigma_0)}\prod_{j=1,\atop j\neq k,j\neq l}^n(1-\Gamma_j\sigma_1)\right].
\end{eqnarray}
To obtain $Z_1(x_0,x_1)$ using Eq.\ (\ref{eqgen}), we express $\prod_{j=1}^p{\rm{sdet}}^{-1/2}\left(X-\Lambda_j\rho\right)$ in terms of the bosonic and fermionic entries of $\rho$ and
expand the resulting expression  in the fermionic variables,
\begin{eqnarray}
&\fl\prod_{j=1}^p{\rm{sdet}}^{-1/2}\left(X-\Lambda_j\rho\right)=\prod_{j=1}^p\frac{x_1-i\Lambda_j\rho_1}{\det^{1/2}(x_0\mathds{1}_2-\Lambda_j\rho_0)}+\sum_{k=1}^p\left[\prod_{j=1,\atop j\neq k}^p\frac{x_1-i\Lambda_j\rho_1}{\det^{1/2}(x_0\mathds{1}_2-\Lambda_j\rho_0)}\right]
\nonumber\\&\fl\frac{\Lambda_k^2}{\det^{3/2}(x_0\mathds{1}_2-\Lambda_k\rho_0)}\left[\eta
\eta^*(x_0-\Lambda_k\rho_{11})+\xi\xi^*(x_0-\Lambda_k\rho_{00})+\eta\xi^*\Lambda_k\rho_{01}-\eta^*\xi\Lambda_k\rho_{01}\right]\\&\fl+\sum_{k,l=1,\atop k\neq l}^p
\left[\prod_{j=1,\atop j\neq k,j\neq l}^p\frac{x_1-i\Lambda_j\rho_1}{\det^{1/2}(x_0\mathds{1}_2-\Lambda_j\rho_0)}\right]\Lambda_k^4\frac{\eta\eta^*\xi\xi^*\left[(x_0-\Lambda_k\rho_{11})(x_0-\Lambda_l\rho_{00})-
\Lambda_k\Lambda_l\rho_{01}^2\right]}{\det^{3/2}(x_0-\Lambda_k\rho_0)\det^{3/2}(x_0-\Lambda_l\rho_0)}.\nonumber
\end{eqnarray}
Inserting this expression together with Eq.\ (\ref{I_1}) into Eq.\ (\ref{eqgen}) we can perform the integrals with respect to the anticommuting variables and the $\rho_1$ integral in the way 
described after Eq.\ (\ref{eq999}). Next we diagonalize the $\sigma_0$ and $\rho_0$ integrals and express it as integrals with respect to their real eigenvalues $s_1$, $s_2$ and $r_1$, $r_2$ and to 
orthogonal $2\times 2$ matrices leading to the functional determinants $|r_1-r_2|$ and $|s_1-s_2|$. A Bessel function is obtained from one integral with respect to the orthogonal matrix. The final 
expression can be written in terms of a fourfold integral with respect to the eigenvalues of $\sigma_0$ and $\rho_0$ and is expressed
in a more compact form by introducing the variables $s=s_1-s_2$, $S=s_1+s_2$, $r=r_1-r_2$ and $R=r_1+r_2$.
\begin{eqnarray}\label{long}
&\fl Z_1(x_0,x_1)=\frac{1}{512\pi}\int\limits_{-\infty}^\infty dS\int\limits_{-\infty}^\infty ds\int\limits_{-\infty}^\infty dR\int\limits_{-\infty}^\infty dr|r||s|\left\{\sum_{u=0}^{p-1}x_1^{p-u}(-1)^uE_u^p(\Lambda)u!J_0\left(\frac{sr}{4}\right)E_u^n(\Gamma)
\right.\nonumber\\&\fl\left.\frac{{\rm e}^{iSR/4}}{
\sqrt{\prod_{j=1}^n\left[1+i\Gamma_j(S+s)/2\right]\left[1+i\Gamma_j(S-s)/2\right]}}-2\sum_{k=1}^n\sum_{l=1}^p\Gamma_k^2\Lambda_l^2\sum_{u=0}^{p-2}x_1^{p-u-1}(-1)^uu!E_u^p(\Lambda^{\hat{l}})\right.\nonumber\\&\fl\left.E_u^p
(\Gamma^{\hat{k}})\frac{1}{\left[x_0-\Lambda_l(R-r)/2\right]}\frac{{\rm e}^{iSR/4}}{\sqrt{
\prod_{j=1}^n\left[1+i\Gamma_j(S+s)/2\right]\left[1+i\Gamma_j(S-s)/2\right]}}    \right.\\&\fl\left.\frac{\left[J_0\left(sr/4\right)-iJ_1\left(sr/4\right)\right]}{\left[1+i\Gamma_k(S-s)/2\right]}
+J_0\left(\frac{sr}{4}\right)\sum_{u=0}^{p-3}x_1^{p-u-2}(-1)^uu!\sum_{m,n=1,\atop m\neq n}^p\sum_{k,l=1,
\atop k\neq l}^n\Gamma_k^2\Gamma_l^2
\right.\nonumber\\&\fl\left.\frac{\Lambda_m^2\Lambda_n^2E_u^n(\Gamma^{\widehat{kl}})E_u^p(\Lambda^{\widehat{mn}}){\rm e}^{iSR/4}}{\left[1+i\Gamma_k(S+s)/2\right]\left[1+i\Gamma_l(S-s)/2\right]
\sqrt{\prod_{j=1}^n\left[1+i\Gamma_j(S+s)/2\right]\left[1+i\Gamma_j(S-s)/2\right]}}
\right.\nonumber\\&\fl\left.\frac{1}{\left[x_0-\Lambda_m(R-r)/2\right]\left[x_0-\Lambda_n(R+r)/2\right]}\right\}  \frac{1}{\sqrt{\prod_{j=1}^p\left[x_0-\Lambda_j(R+r)/2\right]\left[
x_0-\Lambda_j(R-r)/2\right]}}\nonumber
\end{eqnarray}
with the Bessel function of zeroth and of first order $J_0(x)$ and $J_1(x)$, respectively. The eigenvalue density can be calculated from the latter expression via Eq.\
(\ref{specden}) that we split into three terms in order to make the presentation more transparent:
\begin{equation}\label{fulldens}
S_1(x)=S_{1,1}(x)+S_{1,2}(x)+S_{1,3}(x)
\end{equation}
with
\begin{eqnarray}\label{fulldens1}
&\fl S_{1,1}(x)=\frac{1}{512 p\pi}{\rm Im}\left\{\int\limits_{-\infty}^\infty dS\int\limits_{-\infty}^\infty ds\int\limits_{-\infty}^\infty dR\int\limits_{-\infty}^\infty dr|r||s|
\sum_{u=0}^{p-1}x^{p-u-1}(-1)^uE_u^p(\Lambda)u!J_0\left(\frac{sr}{4}\right)
\right.\\ &\fl\left.\frac{E_u^n(\Gamma)(p-u){\rm e}^{iSR/4}}{
\sqrt{\prod\limits_{j=1}^n\left[1+i\Gamma_j(S+s)/2\right]\left[1+i\Gamma_j(S-s)/2\right]}\sqrt{\prod\limits_{j=1}^p\left[x^--\Lambda_j(R+r)/2\right]\left[
x^--\Lambda_j(R-r)/2\right]}}\right\},\nonumber
\end{eqnarray}
\begin{eqnarray}\label{fulldens2}
&\fl S_{1,2}(x)=-\frac{1}{256 p\pi}{\rm Im}\left\{\int\limits_{-\infty}^\infty dS\int\limits_{-\infty}^\infty ds\int\limits_{-\infty}^\infty dR\int\limits_{-\infty}^\infty dr|r||s|\sum_{k=1}^n
\sum_{l=1}^p\Gamma_k^2\Lambda_l^2\sum_{u=0}^{p-2}x^{p-u-2}(-1)^u\right.\nonumber\\&\fl\left.u!E_u^p(\Lambda^{\hat{l}})E_u^p
(\Gamma^{\hat{k}})\frac{(p-u-1)}{\left[1+i\Gamma_k(S-s)/2\right]}\frac{{\rm e}^{iSR/4}}{\sqrt{
\prod_{j=1}^n\left[1+i\Gamma_j(S+s)/2\right]\left[1+i\Gamma_j(S-s)/2\right]}}    \right.\\&\fl\left.\frac{\left[J_0\left(sr/4\right)-iJ_1\left(sr/4\right)\right]}{\left[x^--\Lambda_l(R-r)/2\right]}
\frac{1}{\sqrt{\prod_{j=1}^p\left[x^--\Lambda_j(R+r)/2\right]\left[
x^--\Lambda_j(R-r)/2\right]}}\right\}\nonumber
\end{eqnarray}
and 
\begin{eqnarray}\label{fulldens3}
&\fl S_{1,3}(x)=\frac{1}{512 p\pi}{\rm Im}\left\{\int\limits_{-\infty}^\infty dS\int\limits_{-\infty}^\infty ds\int\limits_{-\infty}^\infty dR\int\limits_{-\infty}^\infty dr|r||s|
J_0\left(\frac{sr}{4}\right)\sum_{u=0}^{p-3}\sum_{m,n=1,\atop m\neq n}^p\sum_{k,l=1,
\atop k\neq l}^n\Gamma_k^2\Gamma_l^2
\right.\\&\fl\left.\frac{x^{p-u-3}(-1)^uu!\Lambda_n^2\Lambda_m^2E_u^n(\Gamma^{\widehat{kl}})E_u^p(\Lambda^{\widehat{mn}}){\rm e}^{iSR/4}}{\left[1+i\Gamma_k(S+s)/2\right]\left[1+i\Gamma_l(S-s)/2\right]
\sqrt{\prod_{j=1}^n\left[1+i\Gamma_j(S+s)/2\right]\left[1+i\Gamma_j(S-s)/2\right]}}
\right.\nonumber\\&\fl\left.\frac{(p-u-2)}{\left[x^--\Lambda_m(R-r)/2\right]\left[x^--\Lambda_n(R+r)/2\right]}\frac{1}{\sqrt{\prod_{j=1}^p\left[x^--\Lambda_j(R+r)/2\right]\left[
x^--\Lambda_j(R-r)/2\right]}}\right\}\nonumber
\end{eqnarray}
 with $x^-=x-i\epsilon$ and $\epsilon\ll1$.
This is the main result of this section. We thus achieved by supersymmetric methods an enormous reduction of the number of integrals: we started from expression (\ref{dens}) containing 
$p\cdot n$ integrals in the real case and end up with only four integrals in Eq.\ (\ref{fulldens}).

It turned out to us to be extremely challenging to bring (\ref{fulldens}) into a form such that the remaining integrals can be calculated numerically. The Bessel functions and the exponential factors
lead to a highly oscillatory behaviour of the integrand. For discussing the evaluation we thus specialize to the case of twofold degeneracies of the eigenvalues of $\Gamma$ and $\Lambda$. Here
the $S$, $R$, $s$ integrals
can be performed analytically at cost of introducing another integral with respect to a finite domain with an integrand monotonously decaying for large $r$. The still remaining two integrals can be 
performed numerically afterwards. We describe the calculation for $S_{1,1}(x)$ in Eq.\ (\ref{fulldens1}). The integral to be analyzed here is given by
\begin{eqnarray}\label{f}
&\fl f(x)={\rm Im}\left\{\int\limits_{-\infty}^\infty dS\int\limits_{-\infty}^\infty ds\int\limits_{-\infty}^\infty dR\int\limits_{-\infty}^\infty dr|r||s|\frac{g(x)J_0(sr/4){\rm e}^{iSR/4}}{
\prod_{j=1}^n\left[1+i\Gamma_j(S+s)/2\right]\left[1+i\Gamma_j(S-s)/2\right]}\right.\nonumber\\& \fl\left.\times\frac{1}{\prod_{j=1}^p\left[x^--\Lambda_j(R+r)/2\right]\left[
x^--\Lambda_j(R-r)/2\right]}\right\},
\end{eqnarray}
where $g(x)$ contains all terms in $S_{1,1}(x)$ that do not depend on the integration variables. The $S$ integral can be performed by residual integration after closing the
contour in the upper half plane for $R\geq0$ and in the lower half plane for $R<0$. For the $r$ dependent terms in the second line of Eq.\ (\ref{f}), we can use relation (\ref{delta00}). As imaginary 
contributions enter into the expression in the curly bracket in the last equation only through an infinitesimal
imaginary part of $x$ and we are finally only interested in imaginary contributions originating from the curly bracket, we can perform the $r$ integral making use of the
$\delta$-distribution to obtain
\begin{eqnarray}
&\fl f(x)=-{\rm Im}\left\{\int\limits_{-\infty}^\infty ds\int\limits_0^\infty dR\frac{|s|}{s}\sum_{k=1}^n\sum_{l=1}^n\frac{32\pi^2g(x)\prod_{j=1}^n\left(-\Gamma^{-2}_j\right)\prod_{j=1}^p\left(-\Lambda^{-2}_j\right)}
{\prod_{j=1,\atop j\neq k}^n\left[(i\Gamma_j)^{-1}-(i\Gamma_k)^{-1}\right]\left[s+(i\Gamma_j)^{-1}-(i\Gamma_k)^{-1}\right]}\right.\nonumber\\ &\fl\left.\times{\rm e}^{isR/4-R/(2\Gamma_k)}\frac{\left|R-2x/\Lambda_l\right|}{\left(R-2x/\Lambda_l\right)}
J_0\left(\frac{s}{4}\left(R-\frac{2x}{\Lambda_l}\right)\right)\prod_{j=1,\atop j\neq l}^p\frac{1}{\left(x/\Lambda_j-x/\Lambda_l\right)\left(R-x/\Lambda_j-x/
\Lambda_l\right)}\right\}.\nonumber\\
\end{eqnarray}
As the remaining expression still contains an oscillatory integrand integrated with respect to an infinite domain we replace the Bessel function by its integral representation and perform the $s$ integral
afterwards using
\begin{eqnarray}
 h(a,b)={\rm Im}\int_0^\infty ds\frac{{\rm e}^{ibs}}{is-a}=\left[{\rm Re}\,{\rm CosI}(iab)-{\rm SinhI}(ab)\right]{\rm e}^{ab}
\end{eqnarray}
with ${\rm CosI}(x)=-\int_x^\infty dt\cos t/t$ and ${\rm SinhI}(x)=\int_0^xdt\sinh t/t$.
We then obtain
\begin{eqnarray}
\displaystyle
 &\fl f(x)=64 \pi \int\limits_0^\infty dR\int\limits_0^\pi d\theta\sum_{k=1}^n\sum_{l=1}^p\frac{\left|R-2x/\Lambda_l\right|g(x){\rm e}^{-R/(2\Gamma_k)}\prod_{j=1}^n\left(\Gamma^{-2}_j\right)\prod_{j=1}^p\left(-\Lambda^{-2}_j\right)}
{\left(R-2x/\Lambda_l\right)\prod_{j=1,\atop j\neq k}^n\left(\Gamma_j^{-1}-\Gamma_k^{-1}\right)}\nonumber\\ &\fl\times
\frac{h\left[\left(\Gamma_k^{-1}-
\Gamma_j^{-1}\right),\left(R/4+\left(R/4-x/2\Lambda_l\right)\cos\theta\right)\right]}{\prod_{j=1\atop j\neq l}^p\left(x/\Lambda_j-x/\Lambda_l\right)
\left(R-x/\Lambda_j-x/\Lambda_l\right)}.
\end{eqnarray}
The integrals in the latter expression are now prepared for performing them numerically. Similar transformations can be done for the other summands in Eq.\ (\ref{fulldens}).

\section{Asymptotic Expressions}\label{sect4}
In this section we calculate the form of the generating function $Z_\beta(x_0,x_1)$ and the eigenvalue density $S_\beta(x)$ in the limit of large $n$ and $p$ with $p/n$ fixed.
\subsection{Limit $n\rightarrow\infty$}
\begin{figure}
\begin{center}
\includegraphics[width=8cm]{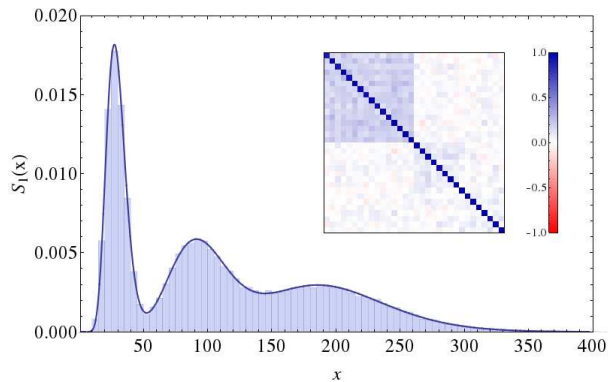}
\caption{We show the eigenvalue density derived from Eq.\ (\ref{assy1}) for $\Lambda_j=1,\, 3.2,\, 6$ (full blue curve) and compare it with a Monte Carlo simulation (histogram). 
The $32\times32$ correlation matrix $D$ is shown in the inset.}
\label{fig10}
\end{center} 
\end{figure}

We first consider $n\rightarrow\infty$ and start from
the expression (\ref{eqgen})  with $I_\beta(\rho)$ from Eq.\ (\ref{Ingha}) inserted, the substitutions $\rho\rightarrow n\rho$ and $\sigma\rightarrow\sigma/n$ yield
\begin{eqnarray}\label{gennn}
\fl Z_\beta(x_0,x_1)=\frac{1}{4}\int d[\rho]d[\sigma]\,{\rm e}^{i\frac{\beta}{2}{\rm str}(\sigma\rho)}\prod_{j=1}^p{\rm sdet}^{-\beta/2}(X-n\rho\Lambda_j)\prod_{j=1}^n
{\rm sdet}^{-\beta/2}(1_{4/\beta}+i\Gamma_j\sigma/n).\nonumber\\
\end{eqnarray}
First we show as in \cite{Wirtz} that in the limit of large $n$ the eigenvalues $\Gamma_j$ in $Z_\beta(x_0,x_1)$ can
be replaced by their geometric mean $\overline{\Gamma}=\sum_{j=1}^n\Gamma_j/n$
\begin{eqnarray}
\label{appprox}
\prod_{j=1}^n{\rm sdet}^{-\beta/2}(\mathds{1}_{4/\beta}+i\Gamma_j\sigma/n)=\exp\left(\frac{\beta}{2}\sum_{m=1}^\infty\frac{(-1)^m}{n^mm}\tr(\Gamma^m){\rm str}\left[(i\sigma)^m\right]\right)\nonumber\\
\approx\exp\left(\frac{n\beta}{2}\sum_{m=1}^\infty\frac{(-1)^m\overline{\Gamma}^m}{n^mm}{\rm str}\left[(i\sigma)^m\right]\right)={\rm sdet}^{-n\beta/2}(\mathds{1}_{4/\beta}+i\overline{\Gamma}\sigma/n).
\end{eqnarray}
In the second step in the last equation we used that the summands with $m>1$ are of lower order in $n$ than the one with $m=1$. We can thus alter the terms with $m>1$ without changing the large $n$
behaviour of the product of superdeterminants in (\ref{appprox}).
Inserting the relation (\ref{appprox}) into (\ref{gennn}), we obtain for $n\rightarrow\infty$
\begin{eqnarray}\label{assy1}
&\fl Z_\beta(x_0,x_1)\sim\frac{1}{4}\!\int \!\!d[\rho]d[\sigma]{\exp}\!\left\{i\frac{\beta}{2}{\rm str}\left[\rho\left(\sigma+i\frac{\mathds{1}_{4/\beta}}{\overline{\Gamma}}\right)\right]\!\right\}
\prod_{j=1}^p{\rm sdet}^{-\beta/2}(X-\rho\Lambda_j){\rm sdet}^{-n\beta/2}\sigma.\nonumber\\ &\fl
\end{eqnarray}
Substituting $\sigma\overline{\Gamma}$ and $\rho/\overline{\Gamma}$ as new integration variables this generating function is the same as the one analyzed in \cite{Rech12} with the replacement 
$\Lambda_j\rightarrow\Lambda_j\overline{\Gamma}$. The eigenvalue density resulting from (\ref{assy1}) can be obtained with the same replacement performed in the eigenvalue density given in 
\cite{Rech12}. To confirm our findings, we compare the analytic expression for the density derived from the generating function~(\ref{assy1}) with numerical simulations. We generate a
sample of $20~000$ real doubly correlated Wishart matrices with $\Lambda=\rm{diag}(6,3.2,1)$ and the $32\times32$ correlation matrix $D$ shown in the inset of Fig.\ \ref{fig10}. As $p/n\ll1$, we obtain   
perfect agreement between our findings and the numerical simulations shown in the main part of Fig.\ \ref{fig10}.

Proceeding analytically, the $\rho$ and $\sigma$ integrals are evaluated by saddle point approximation. The saddle points are determined for the elements of $X$ of order $n$ to be $\sigma_0=-i\mathds{1}_{4/\beta}/\overline{\Gamma}$ and $\rho_0=n\overline{\Gamma}\mathds{1}_{4/\beta}$ yielding
for $Z_\beta(x_0,x_1)$
\begin{equation}
Z_\beta(nx_0,nx_1)\sim\prod_{j=1}^p\frac{x_1-\Lambda_j\overline{\Gamma}}{x_0-\Lambda_j\overline{\Gamma}}.
\end{equation}
This result is for $\overline{\Gamma}=1$ consistent with the one obtained in \cite{Rech12}.
\subsection{Limits $n\rightarrow\infty$ and $p\rightarrow\infty$ with $p/n$ fixed}
Starting again from Eq.\ (\ref{gennn}), $\prod_{j=1}^n{\rm sdet}^{-\beta/2}(\mathds{1}_{4/\beta}+i\Gamma_j\sigma/n)$ can be replaced in the same way as in Eq.\ (\ref{appprox}). Assuming that
$X/n$ is large, a replacement of $\Lambda_j$ by its geometric mean $\overline{\Lambda}$ in $\prod_{j=1}^p
{\rm sdet}^{-\beta/2}(X-n\rho\Lambda_j)$ can be performed in a way analogous to (\ref{appprox}). We then arrive at the following asymptotic expression for
$Z(x_0,x_1)$ for $n\rightarrow\infty$
\begin{eqnarray}\label{genfuc}
\fl Z_\beta(x_0,x_1)\sim\frac{1}{4}\int d[\rho]d[\sigma]{\exp}\left\{{i\frac{\beta}{2}{\rm str}\left[\rho\left(\sigma+i\frac{\mathds{1}_{4/\beta}}{\overline{\Gamma}}\right)\right]}\right\}{\rm sdet}^{-p\beta/2}(X-\rho\overline{\Lambda})
{\rm sdet}^{-n\beta/2}\sigma.\nonumber\\
\end{eqnarray}
The stationary point $(\sigma_0,\rho_0)$ of the latter expression is determined by
\begin{equation}\label{stapha}
\fl n\sigma_0^{-1}=i\rho_0,\hspace*{1cm} \rho_0=\frac{(n-p)\overline{\Gamma}}{2}+\frac{X}{2\overline{\Lambda}}+i\sqrt{\frac{n\overline{\Gamma}X}{\overline{\Lambda}}-
\left(\frac{(n-p)\overline{\Gamma}}{2}+\frac{X}{2\overline{\Lambda}}\right)^2}.
\end{equation}
Performing the integrals in (\ref{genfuc}) by saddle point approximation, we obtain by using Eq.\ (\ref{specden})
for the eigenvalue density the Marchenko-Pastur distribution \cite{Pastur}
\begin{equation}
S_\beta(x)=\frac{n}{2\pi p x\overline{\Lambda}\overline{\Gamma}}{\rm Re}\sqrt{\left[\frac{x}{n}-\left(1-\sqrt{\frac{p}{n}}\right)^2\overline{\Lambda}\overline{\Gamma}\right]
\left[\left(1+\sqrt{\frac{p}{n}}\right)^2\overline{\Lambda}\overline{\Gamma}-\frac{x}{n}\right]}.
\end{equation}

\section{Conclusions}\label{sect5}

We considered data matrices with correlations between time {\it and} position series and calculated the eigenvalue density using supersymmetry. In the unitary case we obtain a closed form expression (without
remaining integrals) that we compare with the result given in Ref.\ \cite{Simon} identifying an error in the expression in the latter article. We compare our result with a Monte Carlo simulation and find
perfect agreement. In the orthogonal case we are at present not aware of any other method like character expansion, Jack or zonal polynomials \cite{Muirhead} to tackle this problem
as the group integrals are much more complicated. Using supersymmetry, we derive an expression for the eigenvalue density in terms of a fourfold integral. Due to the oscillatory character of the integrand it is challenging 
to compute it in the
general case. We restrict to twofold degenerate empirical eigenvalues of the correlation matrices and describe how the fourfold integral can be reduced to a twofold one with a non oscillatory integrand
in that case. Finally we obtain asymptotic expressions for the eigenvalue density in the limit of a large number of time steps $n$  and a large number of
observing points $p$.

It would be interesting in this context to analyze the expression for the eigenvalue density in the orthogonal case further: In the case of nondegenerate eigenvalues an expression that would be numerically
treatable would be highly desirable. Furthermore other quantities characterizing the doubly correlated Wishart ensemble as e.g.\ the distribution of the smallest or largest eigenvalue \cite{Wirtz,
Wirtz0.5,Wirtz1}
or higher order eigenvalue correlation functions could be analyzed.
\section*{Acknowledgements}

We thank the Sonderforschungsbereich Transregio 12 for support. Discussions with Maram Akila and Mario Kieburg are acknowledged.

\end{document}